\documentclass[twocolumn,prd,amsmath,amssymb,floatfix]{revtex4}

\usepackage{bbold}
\usepackage{bm}
\usepackage{color}
\usepackage{dcolumn}
\usepackage{feynmf} 
\usepackage{graphics}
\usepackage{graphicx}
\usepackage{subfigure}
\usepackage{slashed}
\usepackage{placeins}

\allowdisplaybreaks
\usepackage{pdfpages}
\usepackage{eso-pic}
\usepackage{everyshi}
\usepackage{pdflscape}

\usepackage{epsfig}
\usepackage{mathtools}
\usepackage[hidelinks]{hyperref}
\def\al{\alpha}
\def\eps{\epsilon}

\newcommand{\ep}{\varepsilon}

\def\be{\begin{equation}}
\def\ee{\end{equation}}
\def\bea{\begin{eqnarray}}
\def\eea{\end{eqnarray}}
\def\bse{\begin{subequations}}
\def\ese{\end{subequations}}
\def\bc{\begin{center}}
\def\ec{\end{center}}

\def\ra{\rightarrow}

\def\nonum{\nonumber}

\def\I{{\rm i}}

\def\D{{\rm d}}
\def\Ord{{\rm O}}

\newcommand{\ie}{{\it i.e.}}
\newcommand{\eg}{{\it e.g.}}

\begin{document}

\title{Landau-Khalatnikov-Fradkin transformation of the fermion propagator \\ in massless reduced QED}
       \author{A.~James$^{1}$, A.~V.~Kotikov$^{2}$ and S.~Teber$^{1}$}
\affiliation{
$^1$Sorbonne Universit\'e, CNRS, Laboratoire de Physique Th\'eorique et Hautes Energies, LPTHE, F-75005 Paris, France.\\
$^2$Bogoliubov Laboratory of Theoretical Physics, Joint Institute for Nuclear Research, 141980 Dubna, Russia.
 }

\date{\today}

\begin{abstract}
  
  We study the gauge-covariance of the massless fermion propagator in reduced Quantum Electrodynamics (QED).
  Starting from its value in some gauge, we evaluate an all order expression for it in another gauge by means of the Landau-Khalatnikov-Fradkin (LKF) transformation. We find that the weak coupling expansions thus derived are in perfect agreement with the exact calculations. We also prove that the fermion anomalous dimension of reduced QED is gauge invariant to all orders of perturbation theory except for the first one.

\end{abstract}

\maketitle

\section{Introduction}
\label{Sec:Introduction}

The Landau-Khalatnikov-Fradkin (LKF) transformation~\cite{Landau:1955zz} (see also~[\onlinecite{Johnson:1959zz,Sonoda:2000kn}])
is an elegant and powerful transformation allowing one to study the gauge covariance of Green's functions in gauge theories. 
In the latter, gauge freedom is implemented by a covariant gauge fixing procedure that introduces an explicit dependence
of Green's functions on a gauge fixing parameter $\xi$. The LKF transformation then relates the Green's functions in two different $\xi$-gauges. 
Of course, physical quantities should not depend on $\xi$. But important information can be obtained by studying the $\xi$-dependence of various correlation functions.

In its original form, the LKF transformation was applied to the fermion propagator (and also to the fermion-photon vertex that will not be discussed here) 
of four-dimensional quantum electrodynamics (QED$_4$) which is the primary example of an Abelian gauge field theory. 
Since then, it has been extensively used in studies of QED in various dimensions, see, \eg, [\onlinecite{Curtis:1990zs,Burden:1998gr,Jia:2016udu,Fernandez-Rangel:2016zac,
Kizilersu:2009kg,Bashir:2002sp,Jia:2016wyu,Kotikov:2019bqo}] and, more recently, in their generalization to 
brane worlds~\cite{Ahmad:2016dsb} that we shall come back to in the following 
and also to non-abelian $SU(N)$ gauge field theories~\cite{DeMeerleer:2018txc,DeMeerleer:2019kmh}.

As a well known application, let us first mention its crucial role within the study of QED Schwinger-Dyson 
equations see, \eg, [\onlinecite{Curtis:1990zs,Burden:1998gr,Jia:2016udu}], 
where any viable charged-particle-photon vertex ansatz has to satisfy the LKF transformation, both for scalar~\cite{Fernandez-Rangel:2016zac} and 
spinor QED~\cite{Kizilersu:2009kg}. Another notable application~\cite{Bashir:2002sp,Jia:2016wyu} that will be closer to our present concerns
is devoted to estimating the large order behaviour of perturbative expansions. Namely, the non-perturbative nature of the LKF 
transformation fixes certain coefficients appearing in the all-order expansion of the fermion propagator. Given a perturbative propagator 
written for some fixed gauge parameter, say $\eta$, all the coefficients depending on the difference between the gauge fixing parameters of the two
propagators, \ie~  $\xi - \eta$, get fixed by the weak coupling expansion of the LKF-transformed initial propagator. Recently, such
a procedure allowed to prove~\cite{Kotikov:2019bqo} the so-called ``no-$\pi$ theorem''~\cite{Broadhurst:1999xk,Baikov:2010hf,
Baikov:2018wgs,Baikov:2018gap,Baikov:2019zmy}, \eg, 
cancellations involving $\zeta_{2n}$ (or equivalently $\pi^{2n}$) values in the perturbative expansion of Euclidean fermion 
propagator in massless QED$_4$, thereby clarifying the transcendental structure of the latter. 

In the present paper, we apply the LKF transformation to the fermion propagator of massless reduced QED
or RQED$_{d_\gamma, d_{\rm{e}}}$, see Refs.~[\onlinecite{Teber:2012de,Kotikov:2013eha,Teber:2018goo}] and references therein. 
The latter is
 an Abelian gauge theory where the photon and fermion fields live in different space-time dimensionalities, 
 namely the photon is in $d_\gamma$ dimensions whereas the fermion fields are confined to $d_e$ dimensions, 
 where we take $d_e \leq d_\gamma$. We shall focus on the special case of RQED$_{4,3}$ which is an effective
field theory for the so-called planar Dirac liquids, \ie, condensed matter physics systems whose low-energy excitations have a gapless linear, relativistic-like linear dispersion relation and where electrons are confined to a plane ($d_e=2+1$) while interacting via the exchange of photons that can travel through a $d_\gamma=(3+1)$-dimensional bulk.
A prototypical example includes graphene [\onlinecite{PhysRev.71.622,Semenoff:1984dq,Novoselov:2005kj}]. Nowadays, planar Dirac liquids are well observed experimentally 
and are under active study in, \eg, (artificial) graphene-like materials [\onlinecite{polini2013artificial}], surface states of topological
insulators [\onlinecite{RevModPhys.82.3045}], and half-filled fractional quantum Hall systems [\onlinecite{PanCF:2017}]. Interest in RQED$_{4,3}$ also comes
from its connections to $\textrm{QED}_3$~\cite{Kotikov:2016yrn} which is quite often used as an effective field theory of high temperature superconductors [\onlinecite{Marston:1989zz,Ioffe:1989zz,Herbut:2002yq}].

More specifically, we will focus on the case of massless RQED$_{4,3}$. Within the condensed matter context, a vanishing fermion mass implies long-ranged (unscreened) interactions
among the electrons in the absence of doping (the so called intrinsic case). These interactions in turn enforce the flow of the Fermi velocity, \eg, $v \approx c/300$ at experimentally 
accessible scales for graphene, to the velocity of light, $c$, deep in the infra-red (IR) with a corresponding flow of
the fine structure constant, \eg, $\al_g \approx e^2/4\pi \eps \hbar v \approx 2.2$ for graphene, to the usual fine structure constant, $\al_{\rm em} \approx 1/137$. 
Within this context, it is this IR Lorentz invariant fixed point~\cite{Gonzalez:1993uz} that can be described by massless RQED$_{4,3}$~\cite{Teber:2018goo}. A thorough
understanding of this fixed point is a prerequisite to set on a firm ground the study of the physics away from the fixed point which is closer to the experimental reality.  
But this is more difficult to study theoretically, see \eg, [\onlinecite{Teber:2014ita}], the interesting new work [\onlinecite{Munoz-Segovia:2019xip}] and also the recent reviews in Refs.~[\onlinecite{Teber:2016unz,Teber:2019kkp}].

The gauge-covariance of the fermion propagator of massless reduced QED has already been considered in [\onlinecite{Ahmad:2016dsb}]. Here, we carefully 
reconsider this problem using the LKF transformation in the framework of dimensional regularization. We not only 
focus on the bare propagator but also on the renormalized one and provide a detailed comparison between the weak coupling expansion 
of LKF transformed quantities and earlier exact perturbative calculations~\cite{Kotikov:2013eha,Teber:2018goo}.

This paper is organized as follows. In Sec.~\ref{Sec:LKF-reduced-QED}, we start by introducing the position space LKF transformation 
for the general case of reduced QED theories and then derive its momentum space representation for QED$_{4,3}$ that will be the main subject of focus from there on. 
In Sec.~\ref{Sec:LKF-bare-propagator}, a weak coupling expansion of this transformation is performed up to two loops in the $\overline{\textrm{MS}}$-scheme
and its matching with existing perturbative results are discussed. A similar task is carried out in Sec.~\ref{Sec:LKF-renormalization} for the renormalization 
constant and the renormalized propagator. Additionally, we present a proof of the purely one-loop gauge dependence of the fermion anomalous dimension in 
reduced QED. Finally, in Sec.~\ref{Sec:Conclusion}, we summarize our results and conclude. For completeness, various other choices of scales 
are presented in App.~\ref{App:Scales} and in App.~\ref{App:ScalarRQED} the LKF transformation for reduced scalar QED is derived.

\section{LKF transformation for reduced QED}
\label{Sec:LKF-reduced-QED}

We have the following action for reduced $\textrm{QED}_{d_\gamma,d_{\rm e}}$
\begin{flalign}
	S_\textrm{RQED} &= \int d^{d_{\rm e}}x \, \bar{\psi}_\sigma \I D_\mu \gamma^\mu \psi^\sigma \nonumber \\
	&\qquad{}+ \int d^{d_\gamma}x\, \left[-\frac{1}{4} F_{\mu\nu} F^{\mu\nu}  
	- \frac{1}{2\xi} (\partial_\mu A^\mu)^2\right]\, ,
	\label{SRQED}
\end{flalign}
where $\xi$ is the gauge fixing parameter and the flavour index $\sigma$ runs from $1$ to $N_F$. In Eq.~(\ref{SRQED}), the volume elements 
show that the fermion fields $\psi_\sigma$ are confined to $d_{\rm e}$ dimensions whereas the gauge field mediates the interaction through $d_\gamma$ 
space-time dimensions. In explicit form, the (Euclidean space) photon propagator in reduced QEDs reads [\onlinecite{Teber:2012de}]~\footnote{Note that Eq.~(\ref{photon prop}) which
was first derived in [\onlinecite{Teber:2012de}] was also the starting point of the analysis of [\onlinecite{Ahmad:2016dsb}].}:
\begin{align}
	\label{photon prop}
	\widetilde{D}_0^{\mu\nu}(q) = \frac{1}{(4\pi)^{\varepsilon_e}} \frac{\Gamma(1-\varepsilon_e)}{(q^2)^{1-\varepsilon_e}}\left(\delta^{\mu\nu} - (1 - \tilde{\xi})\frac{q^\mu q^\nu}{q^2}\right),
\end{align}
where $\tilde{\xi} = \varepsilon_e + (1 - \varepsilon_e)\xi$ is the \textit{reduced} gauge fixing parameter while we may refer to the original 
gauge fixing parameter $\xi$ as the \textit{bulk} one. In the following, all results will be presented in 
Euclidean space ($d_{\rm \gamma}=4-2\ep_{\rm \gamma}$, $d_{\rm e}=d_{\rm \gamma}-2\ep_{\rm e}$) for QED$_{d_{\rm \gamma},d_{\rm e}}$
 by analogy with the case of QED$_4$.

\subsection{LKF transformation in position space}
\label{Sec:LKF-reduced-QED-x}

We assume that the fermion propagator $S_F(p,\xi)$ in some gauge $\xi$ takes the following general form
\be
S_F(p,\xi) = -\frac{\I}{\hat{p}} \, P(p,\xi) \, ,
\label{SFp}
\ee
%
where $\hat{p}= \gamma^\mu p_\mu$, which contains Dirac $\gamma$-matrices, has been factored out and $P(p,\xi)$ is a scalar function, \ie, its momentum dependence is only via $p^2$.
By analogy, the position-space representation $S_F(x,\xi)$ of the fermion propagator can be written as
\be
S_F(x,\xi) =  \hat{x} \, X(x,\xi)\, ,
\label{SFx}
\ee
where $S_F(x,\xi)$ and $S_F(p,\xi)$ are related to each other with the help of the Fourier transform
\begin{subequations}
\label{SFpx}
\begin{flalign}
	&S_F(p,\xi) = \int \frac{\D^dx}{(2\pi)^{d/2} } \, e^{\I px} \, S_F(x,\xi) \, , \\
	&S_F(x,\xi) = \int \frac{\D^dp}{(2\pi)^{d/2} } \, e^{-\I px} \, S_F(p,\xi) \, .
\end{flalign}
\end{subequations}

In position space, the LKF transformation~\cite{Landau:1955zz,Johnson:1959zz} connects in a very simple way the representations of fermion propagators 
written for different gauge parameters $\xi$  and $\eta$. In dimensional regularization, it takes the following form
\be
S_F(x,\xi) = S_F(x,\eta) e^{D(x)-D(0)}\, ,
\label{LKF}
\ee
where~\cite{Ahmad:2016dsb}
\be
D(x)= f(\ep_{\rm e}) \Delta e^2 \mu^{4-d_{\rm \gamma}} \int \frac{\D^{d_{\rm e}} p}{(2\pi)^{d_{\rm e}} } \, 
\frac{e^{-\I px}}{(p^2)^{2-\ep_{\rm e}}},~~~ \Delta = \xi-\eta\, , 
\label{Dx}
\ee
and the prefactor is given by
\be
f(\ep_{\rm e}) = \frac{\Gamma(2-\ep_{\rm e})}{(4\pi)^{\ep_{\rm e}}} \, , 
\label{fe}
\ee
and follows from the longitudinal part of the photon propagator in Eq.~\eqref{photon prop} above.
%
%
As in the case of QED$_4$, see [\onlinecite{Kotikov:2019bqo}], $D(0)$ is proportional to the massless tadpole and therefore vanishes in dimensional regularization.
%
%
 Hence, Eq.~(\ref{LKF}) takes the simpler form
\be
S_F(x,\xi) = S_F(x,\eta) e^{D(x)}\, ,
\label{LKFN}
\ee
and the remaining task is to compute $D(x)$. This can be achieved using the following simple formulas
for the Fourier transform of massless propagators (see, for example Ref.~[\onlinecite{Kotikov:2018wxe}]):
\begin{subequations}
\label{SFpxN}
\begin{flalign}
&\int \D^dx \, \frac{e^{\I px}}{x^{2\alpha}} \, = \frac{2^{2\tilde{\alpha}} \pi^{d/2} a(\alpha)}{p^{2\tilde{\alpha}}},~~
  a(\alpha) = \frac{\Gamma(\tilde{\alpha})}{\Gamma(\alpha)},~~\tilde{\alpha}=\frac{d}{2}-\alpha\, ,  \\
&\int \D^dp \, \frac{e^{-\I px}}{p^{2\alpha}} \, = \frac{2^{2\tilde{\alpha}} \pi^{d/2} a(\alpha)}{x^{2\tilde{\alpha}}}\, .
\end{flalign}
\end{subequations}
We would like to note that the use of the Euclidean metric simplifies the Fourier transforms thereby illuminating
      the appearance of additional factors such as $\I^k$, where $\I$ is the imaginary unit and the factor $k$ is $\ep$-independent.

So, for  RQED$_{d_{\rm \gamma},d_{\rm e}}$, we have:
\begin{flalign}\label{DxN.0}
D(x) &= \Delta e^2 (\mu^2 x^2)^{2-d_{\rm \gamma}/2} \, \frac{\Gamma(d_{\rm \gamma}/2-2)}{2^4(\pi)^{d_{\rm \gamma}/2}}
\nonum \\
&= \Delta \, A \, \Gamma(d_{\rm \gamma}/2-2) (\pi\mu^2 x^2)^{2-d_{\rm \gamma}/2},
\end{flalign}
where $\displaystyle A = \frac{\alpha_{\rm em}}{4\pi}=\frac{e^2}{(4\pi)^2}$.

Making the dependence on the parameter $\varepsilon$ explicit 
(here and below we shall set $\ep_{\rm \gamma} \equiv \ep$ and  $d_{\rm \gamma} \equiv d$) we finally arrive at the expression
\be
D(x)=-\frac{\Delta A}{\ep} \Gamma(1-\ep)(\pi\mu^2 x^2)^{\ep} \, .
\label{DxN}
\ee
Remarkably, the parameter $\ep_{\rm e}$ has completely disappeared and Eq.~(\ref{DxN}) has exactly the same form as in QED$_4$
with a common factor $\Delta A$, accompanied by the singularity $\ep^{-1}$, contributing to $D(x)$.

Hereafter, we shall only consider the case $d_{\rm e}=3$, \ie, RQED$_{d_{\rm \gamma},3}$, which corresponds (as $d_{\rm \gamma} \to 4$)
to the ultra relativistic limit of graphene (see Ref.~[\onlinecite{Teber:2012de}]). Note that, as it was shown in~[\onlinecite{Teber:2014ita}],
an application of dimensional regularization is very convenient in the non-relativistic limit as well, \ie, where the particles interact via the (instantaneous) Coulomb interaction.

\subsection{LKF transformation in momentum space}
\label{Sec:LKF-reduced-QED-p}

Let $S_F(p,\eta)$, the fermion propagator for some gauge parameter $\eta$ and external momentum $p$, 
take the form (\ref{SFp}) with $P(p,\eta)$ having an expansion
\be
P(p, \eta) = \sum_{m=0}^{\infty} a_m(\eta) A^m {\left(\frac{\tilde{\mu}^2}{p^2}\right)}^{m\ep} \, ,
\label{Peta}
\ee
which is appropriate for the massless case relevant to the present study (as explained in the Introduction, it corresponds to the ultrarelativistic limit of planar Dirac liquids)~\footnote{
	We would also like to note that, in the framework of the three- and four-dimensional QED, the cases with the massive fermion propagators have been considered in Ref.~[\onlinecite{Bashir:2002sp}].}.
 In Eq.~(\ref{Peta}) the $a_m(\eta)$ are coefficients of the loop expansion and $\tilde{\mu}$  the renormalization scale
\be
\tilde{\mu}^2 = 4\pi \mu^2 \, ,
\label{Aem}
\ee
which lies somewhere between the MS scale $\mu$ and $\overline{\rm{MS}}$ scale $\overline{\mu}$.
Then, the LKF transformation shows that for another gauge parameter $\xi$
the result has the form
\be
P(p, \xi) = \sum_{m=0}^{\infty} a_m(\xi) A^m {\left(\frac{\tilde{\mu}^2}{p^2}\right)}^{m\ep} \, ,
\label{Pxi}
\ee
where now
\begin{flalign}
a_m(\xi) &= a_m(\eta) \frac{\Gamma(3/2-(m+1)\ep)}{\Gamma(1+m\ep)}\nonumber \\
&\quad{}\times \sum_{l=0}^{\infty} 
\frac{\Gamma(1+(m+l)\ep)\Gamma^l(1-\ep)}{l!\Gamma(3/2-(m+l+1)\ep)} \, \frac{(\Delta \, A)^l}{(-\ep)^l} \, {\left(\frac{\tilde{\mu}^2}{p^2}\right)}^{l\ep}.
\label{axi}
\end{flalign}
In order to derive Eq.~(\ref{axi}), we used the fermion propagator $S_F(p,\eta)$ with  $P(p,\eta)$ given by Eq.~(\ref{Peta}), 
took the inverse Fourier transform to $S_F(x,\eta)$ and applied the LKF transformation (\ref{LKFN}) in position space. 
As a final step, we took the Fourier transform back to momentum space and obtained $S_F(p,\xi)$ with $P(p,\xi)$ in (\ref{Pxi}).
Let us also note that expansions similar to (\ref{Peta}) and (\ref{axi}) can also be expressed in Minkowski space
with the help of the replacement $p^2 \to - p^2$.

\subsection{$\overline{\rm{MS}}$ scheme}
\label{Sec:LKF-reduced-QED-MSb}

Now let us focus on the 
$\overline{\rm{MS}}$ scale $\overline{\mu}$, which is equal (in the most standard definition) to
\be
\overline{\mu}^2 = \tilde{\mu}^2 e^{-\gamma_E}\, ,
\label{MSbar}
\ee
where $\gamma_E$ is the Euler-Mascheroni constant. As is well known, the $\overline{\rm{MS}}$ scale completely subtracts out the universal factors 
of $\gamma_E$ from the $\ep$-expansions.

In the $\overline{\rm{MS}}$-scheme we can rewrite the result (\ref{axi}) in the following form
\begin{flalign}
a_m(\xi) &= a_m(\eta)
\sum_{l=0}^{\infty} \, \frac{(1-2(m+1)\ep)}{(1-2(m+l+1)\ep)}
\nonumber\\
&\qquad{}\times\tilde{\Phi}(m,l,\ep)
\, \frac{(\Delta \, A)^l}{(-\ep)^l l!} \, {\left(\frac{\overline{\mu}^2}{p^2}\right)}^{l\ep} \, ,
\label{axi.1}
\end{flalign}
where we have purposefully extracted the factor $(1-2(m+1)\ep)/(1-2(m+l+1)\ep)$ from $\tilde{\Phi}(m,l,\ep)$ in order to have \textit{equal transcendental level}, \ie, the same
$s$ values of $\zeta_s$ in the $\ep$-expansion of $\tilde{\Phi}(m,l,\ep)$ (see below). As will be shown below, the factor $\tilde{\Phi}(m,l,\ep)$ reading
\be
\tilde{\Phi}(m,l,\ep)=\frac{\Gamma(1/2-(m+1)\ep)\Gamma(1+(m+l)\ep)\Gamma^{l}(1-\ep)}{
  \Gamma(1+m\ep)\Gamma(1/2-(m+l+1)\ep)e^{-l\gamma_E\ep}} \, ,
\label{Phi.V.1}
\ee
can be written as an expansion in the $\zeta_i$ $(i\geq 2)$ Euler constants. Note that the $\gamma_E$-dependent term 
arises from the redefinition (\ref{MSbar}) of the scale $\tilde{\mu} \to \overline{\mu}$.

At this point, it is convenient to re-express the $\Gamma$-functions with arguments close to half-integer ones 
using the standard property (Legendre duplication formula):
\be
\Gamma(2\alpha)= \frac{2^{2\alpha-1}}{\Gamma(1/2)} \, \Gamma(\alpha) \Gamma(\alpha + 1/2) \, ,
\label{Gamma.1}
\ee
which leads to the following relation
\begin{flalign}
\frac{\Gamma(1/2-(m+1)\ep)}{\Gamma(1/2-(m+l+1)\ep)} &= \frac{1}{2^{2l\ep}} \,
\frac{\Gamma(1-2(m+1)\ep)}{\Gamma(1-(m+1)\ep)} \nonumber\\
&\qquad{}\times\frac{\Gamma(1-(m+l+1)\ep)}{\Gamma(1-2(m+l+1)\ep)}\, .
\label{Property}
\end{flalign}
Then, we may write
\be
\tilde{\Phi}(m,l,\ep)= \frac{1}{4^{l\ep}} \, \Phi(m,l,\ep)\, ,
\label{Phi.V.2}
\ee
with
\begin{flalign}
\Phi(m,l,\ep) &= \frac{\Gamma(1-2(m+1)\ep)\Gamma(1+(m+l)\ep)}{\Gamma(1-(m+1)\ep)
  \Gamma(1+m\ep)}\nonumber\\
&\quad \times\frac{\Gamma(1-(m+l+1)\ep)\Gamma^{l}(1-\ep)}{\Gamma(1-2(m+l+1)\ep)e^{-l\gamma_E\ep}}\, ,
\label{Phi}
\end{flalign}
and Eq.~(\ref{axi.1}) can be represented as
\be
a_m(\xi) = a_m(\eta)
\sum_{l=0}^{\infty} \,
\overline{\Phi}(m,l,\ep)
\, \frac{(\Delta \, A)^l}{(-\ep)^l l!} \, {\left(\frac{\overline{\mu}^2}{4p^2}\right)}^{l\ep} \, ,
\label{axi.2}
\ee
with
\be
\overline{\Phi}(m,l,\ep) = \frac{1-2(m+1)\ep}{1-2(m+l+1)\ep} \, \Phi(m,l,\ep) \, .
\label{OPhi}
\ee
We would like to draw the attention of the reader to the redefinition of the argument in the r.h.s.\ of (\ref{axi.2}):
$\overline{\mu}^2/p^2 \to \overline{\mu}^2/(4p^2)$. Such a redefinition amounts to subtracting factors of $\ln 2$. As we shall see below, 
it agrees with the exact perturbative calculations done with the $\overline{\textrm{MS}}$ scale (see
[\onlinecite{Teber:2018goo}]). Note that the latter include an additional negative sign for momentum squared in the denominator 
because the results of [\onlinecite{Teber:2018goo}] were given in Minkowski space. Therefore, $p^2_E = -p^2_M$ under Wick rotations, 
in the mostly minus signature that was used in that paper.

\subsection{$\ep$-expansion}
\label{Sec:LKF-reduced-QED-exp}

Let us recall that the $\Gamma$-function $\Gamma(1+\beta\ep)$ has the following expansion around $1$:
\be
\Gamma(1+\beta\ep) = \exp \Big[ -\gamma_E \beta \ep + \sum_{s=2}^{\infty}\, (-1)^s \, \eta_s \beta^s \ep^s \Bigr], \quad 
\eta_s = \frac{\zeta_s}{s} \, .
\label{Gamma.2}
\ee
So, the factor $\Phi(m,l,\ep)$ can be written as:
\be
\Phi(m,l,\ep)= \exp \Big[ \sum_{s=2}^{\infty}\,\eta_s \, p_s(m,l) \, \ep^s \Bigr]\, ,
\label{Phi.MSbar}
\ee
where (but now including $s \geq 1$)
\begin{flalign}
p_s(m,l)&= (2^s-1) \, \Bigl\{(m+1)^s-(m+l+1)^s \Bigr\} \nonumber\\
&\qquad{} + (1+\delta_s^1) l +(-1)^s \Bigl\{(m+l)^s-m^s\Bigr\}\, ,
\label{p.s}
\end{flalign}
($\delta_s^1$ is the Kronecker symbol) and, indeed,
\be
p_1(m,l)=0,~~ p_2(m,l)=-l\Big(4m+5+2l\Bigr)\, ,
\label{p.2}
\ee
\ie, the $\overline{\rm{MS}}$-scale takes out the Euler constant $\gamma_E$ from consideration. 

As can be seen from Eq.~(\ref{Phi.MSbar}), the factor $\Phi(m,l,\ep)$ contains $\zeta_s$-values with the same weight $s$ in front of $\ep^s$.
This is rather similar to what was found in Ref.~[\onlinecite{Kotikov:2000pm}]. In some cases, such a property 
allows to derive results without any detailed calculations (as in Ref.~[\onlinecite{Kotikov:2002ab}]).
In other cases, it simplifies the structure of the results, which can be predicted as an ansatz in a very simple way
(see Refs.~[\onlinecite{Fleischer:1998nb,Kotikov:2007cy}]). For recent applications of this property to QCD and super Yang-Mills, 
see the papers~[\onlinecite{Dixon:2019uzg}] and references and discussions therein. 

Recently, this property was also applied to the LKF transformation of QED$_4$ in [\onlinecite{Kotikov:2019bqo}] by some of the present authors.
Combined with an appropriate choice of scale, it led to an all-order proof~\cite{Kotikov:2019bqo} that the perturbative series can be 
exactly expressed in terms of a hatted transcendental basis that eliminates all even $\zeta$-values, \ie, the no-$\pi$ theorem~\cite{Broadhurst:1999xk,Baikov:2010hf,
Baikov:2018wgs,Baikov:2018gap,Baikov:2019zmy}.
In the case of QED$_{4,3}$, the situation is not so simple. As can be seen from Eq.~(\ref{p.2}), the fact that $p_2(m,l) \neq 0$ means that 
$\zeta_2$-values cannot be subtracted out, unlike in the even-dimensional case~\cite{Kotikov:2019bqo}. 
As shown in App.~\ref{App:Scales}, other choices of scale are possible but do not further simplify the transcendental structure of 
this (partially) odd dimensional theory, see also Ref.~[\onlinecite{Broadhurst:1996yc}] for an early study.

Thus, in this section, we have obtained a series representation, Eq.~\eqref{axi.2}, for the LKF transformation of the fermion propagator of reduced QED$_{4,3}$ 
in the $\overline{\textrm{MS}}$-scheme (see App.~\ref{App:Scales} for other choices of scales and App.~\ref{App:ScalarRQED} for an analogous expression 
in the case of reduced scalar QED). We now need to verify that 
the gauge dependence produced by this transformation agrees with exact perturbative results (known in the literature up to the 2-loop order). The next two
sections are devoted to this task.

\section{LKF transformation for the bare fermion propagator}
\label{Sec:LKF-bare-propagator}

\subsection{Bare fermion propagator}
\label{Sec:LKF-bare-propagator:PT}

The calculations of the photon and fermion propagators in the framework of the reduced QED have been done in Refs.~[\onlinecite{Teber:2012de}]
and [\onlinecite{Kotikov:2013eha,Teber:2018goo}], respectively (see also the recent reviews in [\onlinecite{Teber:2016unz,Teber:2019kkp}]).

The fermion propagator (\ref{SFp}) can be represented in the following form
\be
P(p, \xi)=\frac{1}{1-\Sigma_V(p^2,\xi)}\, ,
\label{Pp}
\ee
where the fermion self-energy $\Sigma_V(p^2,\xi)$ can be written with next-to-leading order (NLO) accuracy as
\be
\Sigma_V(p^2,\xi)= \Sigma_{1 V}(p^2,\xi) + \Sigma_{2 V}(p^2,\xi)\, .
\label{Sigmav}
\ee
Here $\Sigma_{1 V}(p^2,\xi)$ and $\Sigma_{2 V}(p^2,\xi)$ are the one-loop and two-loop contributions to the self-energy.
Their bare contributions can be represented in the following simple form:
\footnote{Since here we are working in Euclidean space, all the factors $(\overline{\mu}^{2}/(-p^{2}))$ in Ref.
\cite{Teber:2018goo} should be replaced by   $(\overline{\mu}^{2}/p^{2})$.}

\be
\Sigma_{l V}(p^2,\xi) = A^l \, \overline{\Sigma}_{l V}(\xi)\, {\left(\frac{\overline{\mu}^{2}}{4p^{2}}\right)}^{l\ep}\, ,
\label{Sigma}
\ee
where the coefficients $\overline{\Sigma}_{l V}(\xi)$ are now just expressions without $\bar{\mu}$- or $p$-dependence anymore.

The one-loop term $\overline{\Sigma}_{1 V}(\xi)$ takes the following form~\cite{Teber:2018goo}
\begin{flalign}
\overline{\Sigma}_{1 V}(\xi) &= \frac{1-3\xi}{3\ep} + \frac{10}{9} - 2\xi \nonumber\\
&\qquad{} + \left(\frac{112}{27} - 8\xi -\frac{7(1-3\xi)}{6}\zeta_2\right) \ep
+ O(\ep^2) \, .
\label{Sigma1}
\end{flalign}
The two-loop term, $\overline{\Sigma}_{2 V}(\xi)$, can be represented as a sum of three contributions corresponding to three distinct Feynman diagrams~\cite{Teber:2018goo}
\be
\overline{\Sigma}_{2 V}(\xi) = \overline{\Sigma}_{2a V}(\xi) +  \overline{\Sigma}_{2b V}(\xi)  + \overline{\Sigma}_{2c V}(\xi)\, ,
\label{Sigma2}
\ee
with
\begin{subequations}
\label{Sigma2.1}
\begin{flalign}
\overline{\Sigma}_{2a V}(\xi) &= - 4N_F \zeta_2 \left(\frac{1}{\ep} + 2\ln 4\right) + O(\ep) \, ,  \\
\overline{\Sigma}_{2b V}(\xi) &= \frac{(1-3\xi)^2}{18\ep^2} + \left(\frac{11}{27}  -\frac{7\xi}{9}\right)\, \frac{1-3\xi}{\ep}
+ \frac{206}{81}\nonumber\\
&\qquad{} + 2\,(7\xi-6)\xi -\frac{(1-3\xi)^2}{2}\zeta_2 + O(\ep) \, ,  \\
\overline{\Sigma}_{2c V}(\xi) &= -\frac{(1-3\xi)^2}{9\ep^2} + \left(-\frac{37}{27} +\frac{(34-39\xi)\xi}{9}\right)\, \frac{1}{\ep}
\nonumber \\
&\hspace{-1cm}- \frac{1390}{81} + \frac{532\xi}{27} -22\xi^2 + \frac{71+21\xi(3\xi-2)}{9}\zeta_2 + O(\ep) \, .
\end{flalign}
\end{subequations}
We note that the part $\overline{\Sigma}_{2a V}(\xi) $ is $\xi$-independent and, thus, the full result can be represented in the form
\be
\overline{\Sigma}_{2 V}(\xi) = \overline{\Sigma}_{2a V} +  \overline{\Sigma}_{2bc V}(\xi)\, ,
\label{Sigma2.2}
\ee
where the contribution
\be
\overline{\Sigma}_{2bc V}(\xi)=  \overline{\Sigma}_{2b V}(\xi)  + \overline{\Sigma}_{2c V}(\xi)\, ,
\label{Sigma2bc}
\ee
has the following expression
\begin{flalign}
\overline{\Sigma}_{2bc V}(\xi) &= -\frac{(1-3\xi)^2}{18\ep^2} - \frac{2}{27\ep} \, \Bigl(13 -3\xi(8-9\xi)\Bigr) \nonumber\\
&\quad{}-  \frac{8}{81} \, \Bigl(148 -3\xi(26-27\xi)\Bigr) \nonumber\\
&\quad{}+  \frac{\zeta_2}{18} \, \Bigl(128 +5(1-3\xi)^2\Bigr) + O(\ep) \, .
\label{Sigma2.3}
\end{flalign}

As for the propagator itself, at the NLO approximation, Eq.~(\ref{Pp}) can be rewritten as
%
\be
P(p,\xi)=1+\Sigma_{1 V}(p^2,\xi) + \Sigma^2_{1 V}(p^2,\xi) + \Sigma_{2 V}(p^2,\xi) + \cdots \, ,
\label{Pp.1}
\ee
where $\Sigma_{1 V}(p^2)$ has the form (\ref{Sigma}) with $l = 1$ and the contribution $\Sigma^2_{1 V}(p^2) + \Sigma_{2 V}(p^2)$
can be represented as
\be
\Sigma^2_{1 V}(p^2,\xi) + \Sigma_{2 V}(p^2,\xi) = A^2 \Bigl(\overline{\Sigma}_{2a V} + \widetilde{\Sigma}_{2bc V}(\xi) \Bigr)\,{\left(\frac{\mu^{2}}{4p^{2}}\right)}^{2\ep}\, ,
\label{Sigma.2}
\ee
with
\begin{flalign}
\widetilde{\Sigma}_{2bc V}(\xi) &= \overline{\Sigma}_{2bc V}(\xi) + \overline{\Sigma}^2_{1 V}(\xi)\nonumber\\
&= \frac{(1-3\xi)^2}{18\ep^2} - \frac{1}{\ep} \, \left( \frac{2}{9} + \frac{16}{9}\xi -2\xi^2 \right)
\nonumber \\ &\quad{}
+  \zeta_2 \, \left( \frac{64}{9} - \frac{1}{2}(1-3\xi)^2 \right)
\nonumber\\
&\quad{}
-  \frac{4}{81} \, \Bigl(215 +3\xi(70-81\xi)\Bigr) + O(\ep) \, .
\label{Sigma2.4}
\end{flalign}
Let us note that the last term $\widetilde{\Sigma}_{2bc V}(\xi)$ contains all the $\xi$-dependence at the NLO level of accuracy.

\subsection{LKF transformation
}
\label{Sec:LKF-bare-propagator:LKF}

With the help of the results of Secs.~\ref{Sec:LKF-reduced-QED-MSb} and \ref{Sec:LKF-bare-propagator:PT} above,
we can deduce that the one- and two-loop results for the fermion propagator
in two different gauges are related to each other in the following way:
\begin{subequations}
\label{OSigma.1+2}
\begin{flalign}
&\overline{\Sigma}_{1 V}(\xi)= \overline{\Sigma}_{1 V}(\eta) +  \overline{\Sigma}_{0 V}(\eta) \, \overline{\Phi}(0,1,\ep) \, \frac{\Delta}{(-\ep)}\, ,
\label{OSigma.1} \\
&\overline{\Sigma}_{2 V}(\xi)+\overline{\Sigma}^2_{1 V}(\xi)= \overline{\Sigma}_{2 V}(\eta) + \overline{\Sigma}^2_{1 V}(\eta)  
\nonum \\
&\hspace{0.2cm} +  
\overline{\Sigma}_{1 V}(\eta) \, \overline{\Phi}(1,1,\ep) \, \frac{\Delta}{(-\ep)}  + \overline{\Sigma}_{0 V}(\eta) \, \overline{\Phi}(0,2,\ep) \, \frac{\Delta^2}{2(-\ep)^2}\, ,
\label{OSigma.2}
\end{flalign}
\end{subequations}
with $\overline{\Sigma}_{0 V}(\eta)=1$.

Taking $\eta = 0$, \ie,  starting from the Landau gauge and the fact that the contribution $\overline{\Sigma}_{2a V}$ is gauge invariant,
we have that
\begin{subequations}
\label{OSigma.1a+2a}
\begin{flalign}
&\overline{\Sigma}_{1 V}(\xi)= \overline{\Sigma}_{1 V}(\xi=0) +   \overline{\Phi}(0,1,\ep) \, \frac{\xi}{(-\ep)}\, ,
\label{OSigma.1a} \\
&\widetilde{\Sigma}_{2bc V}(\xi)= \widetilde{\Sigma}_{2bc V}(\xi=0) +  \overline{\Sigma}_{1 V}(\xi=0) \, \overline{\Phi}(1,1,\ep) \, \frac{\xi}{(-\ep)}\nonumber\\
&\hspace{2cm}
+ \overline{\Phi}(0,2,\ep) \, \frac{\xi^2}{2(-\ep)^2}\, ,
\label{OSigma.2a}
\end{flalign}
\end{subequations}
where the results for $\overline{\Sigma}_{1 V}(\xi=0)$ and $\widetilde{\Sigma}_{2bc V}(\xi=0)$ can be obtained from Eqs.~(\ref{Sigma1}) and
(\ref{Sigma2.4})  after setting $\xi = 0$. This yields:
\begin{subequations}
\label{Sigma1a+Sigma2.4a}
\begin{flalign}
&\overline{\Sigma}_{1 V}(\xi=0) = \frac{1}{3\ep} + \frac{10}{9} + \left(\frac{112}{27} -\frac{7}{6}\zeta_2\right) \ep
+ O(\ep^2) \, , \label{Sigma1a} \\
&\widetilde{\Sigma}_{2bc V}(\xi=0) =  \frac{1}{18\ep^2} - \frac{2}{9\ep} + \frac{119}{18} \zeta_2
-  \frac{860}{81} + O(\ep) \, .
\label{Sigma2.4a}
\end{flalign}
\end{subequations}
With the help of Eqs.~(\ref{OPhi}), (\ref{Phi.MSbar}) and (\ref{p.s}), we find that the expansions of $\overline{\Phi}(m,l,\ep)$
for the cases of interest read:
\begin{subequations}
\label{Phi-exp}
\begin{flalign}
&\overline{\Phi}(0,1,\ep)=1+2\ep+ \left(8-\frac{7}{2} \zeta_2\right)\ep^2, \, \\
&\overline{\Phi}(1,1,\ep)=1+2\ep+ \left(12-\frac{11}{2} \zeta_2\right)\ep^2 \, ,  \\
&\overline{\Phi}(0,2,\ep)=1+4\ep+ \Bigl(24-9 \zeta_2\Bigr)\ep^2 \, .
\end{flalign}
\end{subequations}
Then Eqs.~(\ref{OSigma.1a+2a}), together with the expressions of $\overline{\Sigma}_{1 V}(\xi=0)$ and $\widetilde{\Sigma}_{2bc V}(\xi=0)$ in (\ref{Sigma1a}) and (\ref{Sigma2.4a}) 
as well as the $\ep$-expansions of Eqs.~(\ref{Phi-exp}) immediately allow to reproduce the full results for $\overline{\Sigma}_{1 V}(\xi)$ and $\widetilde{\Sigma}_{2bc V}(\xi)$ 
presented in the previous sections, Eqs.~(\ref{Sigma1}) and (\ref{Sigma2.4}).

Thus, we have verified that the bare results for $\overline{\Sigma}_{1 V}(\xi)$ and $\widetilde{\Sigma}_{2bc V}(\xi)$ are exactly in agreement with the LKF transformation
(using dimensional regularization, our derivations proceed without any replacements involving a cut-off parameter $\Lambda$ and the scale $\mu$ as in the case of 
Ref.~[\onlinecite{Ahmad:2016dsb}]).

\section{LKF transformation and renormalization}
\label{Sec:LKF-renormalization}

\subsection{Renormalized fermion propagator in momentum space}
\label{Sec:LKF-renormalization:PT}

Since all renormalized results are constructed from the bare ones through the Bogoliubov-Parasiuk-Hepp-Zimmermann (BPHZ) procedure 
(for a definition of the procedure, see for example,  Refs.~[\onlinecite{Teber:2018goo,Teber:2019kkp}]), 
all results including the renormalized ones must be in agreement with the LKF transformation, too. 

In order to show this explicitly to the 2-loop order, let us first note that the fermion propagator given by Eq.~(\ref{SFp}) is the unrenormalized one. It can be
conveniently factored as:
\be
P(p,\xi) = Z_\psi(A, \xi)\,P_r(p,\xi)\, ,
\label{SFp-ren}
\ee
where we have taken into account the fact that $A$ and $\xi$ are \textit{not renormalized} in QED$_{4,3}$, \ie, $A_r \equiv A$ and
$\xi_r \equiv \xi$. In Eq.~(\ref{SFp-ren}), the renormalization constant $Z_\psi(A, \xi)$ and the renormalized fermion propagator $P_r(p,\xi)$
can be expanded as:
\begin{subequations}
\label{Z+Pr:loop-expansion}
\begin{flalign}
&Z_\psi(A,\xi) = 1\! +\! \!\sum_{l=1}^{+\infty} Z_{l \psi}(\xi)\,A^l;   Z_{l \psi}(\xi) = \!\sum_{j=-l}^{-1} Z_{\psi}^{(l,j)}(\xi)\,\ep^j\, ,
 \label{Zpsi:loop-expansion} \\
&P_r(p,\xi) = 1\! +\! \!\sum_{l=1}^{+\infty} P_{lr}(p,\xi)\,A^l;   P_{lr}(p,\xi) = \!\sum_{j=0}^{+\infty} P_r^{(l,j)}(p,\xi)\,\ep^j.
 \label{Pr:loop-expansion}
\end{flalign}
\end{subequations}
The renormalization constant and renormalized propagator have been computed~\cite{Teber:2018goo} up to two loops in reduced QED, for arbitrary $\ep$ at one-loop
and to $\Ord(\ep^0)$ for the propagator. The one-loop expressions read~\cite{Teber:2018goo}:
\begin{subequations}
\label{Z+Pr:1loop-expression}
\begin{flalign}
	&Z_{1\psi}(\xi) = \frac{1-3\xi}{3\ep}\, ,
 \label{Zpsi:1loop-expression} \\
	&P_{1r}(p,\xi) = \frac{10}{9} - \frac{1-3\xi}{3}\,L_p - 2\xi +\nonumber\\
	&\hspace{2cm} + \left( \frac{112}{27} - \frac{7\zeta_2}{6} - \frac{10 L_p}{9} + \frac{L_p^2}{6} - \right . 
	\nonum \\
	&\hspace{2cm} \left . - \xi\,\left( 8 - \frac{7\zeta_2}{2} - 2 L_p + \frac{L_p^2}{2}\right) \right)\,\ep + \Ord(\ep^2)\, ,
 \label{Pr:1loop-exppression}
\end{flalign}
\end{subequations}
where $L_p = \ln(4 p^2 / \overline{\mu}^2)$. The two-loop expressions are given by~\cite{Teber:2018goo}:
\begin{subequations}
\label{Z+Pr:2loop-expression}
\begin{flalign}
	&Z_{2\psi}(\xi) = \frac{(1-3\xi)^2}{18 \ep^2} - \frac{4}{\ep}\,\left( N_F \zeta_2 + \frac{4}{27} \right)\, ,
 \label{Zpsi:2loop-expression} \\
	&P_{2r}(p,\xi) = 8 N_F \zeta_2\,( L_p - 2\ln 2)  - 12 + 7 \zeta_2 + \frac{22}{27}\,L_p +\nonumber\\
	&\hspace{2cm} + \frac{L_p^2}{18} - \frac{\xi}{9}\,\left( 32 - 6 \zeta_2 - 16 L_p + 3 L_p^2 \right ) +
        \nonum \\
	&\hspace{2cm} + \xi^2\,\left( 4 - \zeta_2 - 2 L_p + \frac{L_p^2}{2}\right) + \Ord(\ep)\, .
 \label{Pr:2loop-exppression}
\end{flalign}
\end{subequations}
Let us further note that these expressions allow one to compute the fermion anomalous dimension up to two loops with the help of the relation:
\be
\gamma_\psi(A,\xi) = \sum_{l=1}^{\infty} \gamma_{\psi,l}(\xi)\,A^{l}, \quad \gamma_{\psi,l}(\xi) = 2\,l\,Z_\psi^{(l,-1)}(\xi)\, ,
\label{gammapsi:gen}
\ee
yielding~\cite{Teber:2018goo}:
\be
\gamma_\psi(A,\xi) = 2\,A\,\frac{1-3\xi}{3} - 16\,A^2\,\left( N_F \zeta_2 + \frac{4}{27} \right) + \Ord(A^3)\, .
\label{gammapsi:express}
\ee

\subsection{LKF transformation in momentum space}
\label{Sec:LKF-renormalization:LKF}

We shall now determine the LKF transformations of $Z_{l\psi}(\xi)$ and $P_{lr}(p,\xi)$ up to two-loop and compare the obtained results with those of the last subsection.
In order to proceed, we first note that, at NLO, Eq.~(\ref{SFp-ren}) can be written as:
\begin{flalign}
P(p,\xi) &= 1 + A\,\bigg(Z_{1\psi}(\xi) + P_{1r}(p,\xi) \bigg) +\nonumber\\
&\hspace{-1cm}+ A^2\,\bigg(Z_{2\psi}(\xi) + Z_{1\psi}(\xi)\,P_{1r}(p,\xi) + P_{2r}(p,\xi) \bigg) + \Ord(A^3)\, .
\label{SFp-ren2}
\end{flalign}
Comparing (\ref{SFp-ren2}) with (\ref{Pp.1}) and using the notations of (\ref{Sigma}) then yields:
\begin{subequations}
\label{Z+Pr vs Sigma}
\begin{flalign}
        &Z_{1\psi}(\xi) + P_{1r}(p,\xi) = \overline{\Sigma}_{1 V}(\xi) {\left(\frac{\overline{\mu}^{2}}{4p^{2}}\right)}^{\ep} \, ,
 \label{Z+Pr vs Sigma 1loop} \\
	&Z_{2\psi}(\xi) + Z_{1\psi}(\xi) P_{1r}(p,\xi) + P_{2r}(p,\xi) = \widetilde{\Sigma}_{2 V}(\xi)\, {\left(\frac{\overline{\mu}^{2}}{4p^{2}}\right)}^{2\ep},\nonumber\\
	& \widetilde{\Sigma}_{2 V} = \overline{\Sigma}_{2 V} + \overline{\Sigma}_{1 V}^2\, ,
 \label{Z+Pr vs Sigma 2loop}
\end{flalign}
\end{subequations}
where $\overline{\Sigma}_{l V}(\xi)$ has the following $\ep$-expansion:
\be
\overline{\Sigma}_{l V}(\xi) = \sum_{j=-l}^{+\infty} \overline{\Sigma}_{V}^{(l,j)}(\xi)\,\ep^j\, .
\ee
The LKF transformations of $Z_{l\psi}(\xi)$ and $P_{lr}(p,\xi)$ can then be obtained by identifying identical powers of $\ep$ on both sides of Eqs.~(\ref{Z+Pr vs Sigma}). 

At one loop, this straightforwardly yields:
\begin{subequations}
\label{Z+Pr vs Sigma 1loop xi}
\begin{flalign}
	&Z_{1\psi}(\xi) = Z_{1\psi}(0) - \frac{\xi}{\ep} \, ,
 \label{Z vs Sigma 1loop xi} \\
	&P_{1r}(p,\xi) = P_{1r}(p,0) + (L_p - 2)\, \xi - \nonumber\\
	&\hspace{2cm} - \xi\,\left( 8 - \frac{7\zeta_2}{2} - 2 L_p + \frac{L_p^2}{2} \right)\,\ep + \Ord(\ep^2)\, ,
 \label{Pr vs Sigma 1loop xi}
\end{flalign}
\end{subequations}
where
\begin{subequations}
\label{Z+Pr vs Sigma 1loop xi=0}
\begin{flalign}
        &Z_{1\psi}(0) =  \frac{\overline{\Sigma}_{V}^{(1,-1)}(0)}{\ep} \, ,
 \label{Z vs Sigma 1loop xi=0} \\
	&P_{1r}(p,0) = \overline{\Sigma}_{V}^{(1,0)}(0) - L_p\,\overline{\Sigma}_{V}^{(1,-1)}(0) +\nonumber\\
	&\hspace{0.12cm} + \bigg( \overline{\Sigma}_{V}^{(1,1)}(0) - L_p\,\overline{\Sigma}_{V}^{(1,0)}(0) 
	+ \frac{L_p^2}{2}\,\overline{\Sigma}_{V}^{(1,-1)}(0)\bigg)\,\ep + \Ord(\ep^2)\, .
 \label{Pr vs Sigma 1loop xi=0}
\end{flalign}
\end{subequations}
As for the two loop case, we first note that:
\be
\widetilde{\Sigma}_{2 V}(\xi) = \overline{\Sigma}_{2a V}(\xi) + \widetilde{\Sigma}_{2bc V}(\xi)\, ,
\ee
which in component form can be written as:
\begin{subequations}
\label{tSigma2Vxi}
\begin{flalign}
	&\widetilde{\Sigma}_{V}^{(2,-2)}(\xi) = \widetilde{\Sigma}_{bc V}^{(2,-2)}(\xi) \, ,
 \label{tSigma2Vxi(2,-2)} \\
 &\widetilde{\Sigma}_{V}^{(2,-1)}(\xi) = -4N_F\zeta_2 + \widetilde{\Sigma}_{bc V}^{(2,-1)}(\xi) \, ,
 \label{tSigma2Vxi(2,-1)} \\
 &\widetilde{\Sigma}_{V}^{(2,0)}(\xi) = -16N_F\zeta_2\,\ln 2 +  \widetilde{\Sigma}_{bc V}^{(2,0)}(\xi) \, ,
 \label{tSigma2Vxi(2,0)} 
\end{flalign}
\end{subequations}
where we restricted to $\widetilde{\Sigma}_{V}^{(2,j)}$ with $j\leq 0$. Then, using Eq.~(\ref{OSigma.2a}) yields:
\begin{subequations}
\label{tSigma2bcVxi}
\begin{flalign}
 &\widetilde{\Sigma}_{bcV}^{(2,-2)}(\xi) = \widetilde{\Sigma}_{bc V}^{(2,-2)}(0) + \frac{\xi^2}{2} - \xi\,\overline{\Sigma}_{V}^{(1,-1)}(0)\, ,
 \label{tSigma2bcVxi(2,-2)} \\
 &\widetilde{\Sigma}_{bcV}^{(2,-1)}(\xi) = \widetilde{\Sigma}_{bc V}^{(2,-1)}(0) + 2\xi^2 - \xi\,\overline{\Sigma}_{V}^{(1,0)}(0)\! - \! 2\xi\,\overline{\Sigma}_{V}^{(1,-1)}(0)\, ,
 \label{tSigma2bcVxi(2,-1)} \\
 &\widetilde{\Sigma}_{bcV}^{(2,0)}(\xi) = \widetilde{\Sigma}_{bc V}^{(2,0)}(0) + \xi^2\,\left( 12 - \frac{9\zeta_2}{2}\right) - 
 \xi\,\overline{\Sigma}_{V}^{(1,1)}(0)- 
 \nonum \\
	&\hspace{2cm} - 2\xi\,\overline{\Sigma}_{V}^{(1,0)}(0) - \xi\,\left( 12 - \frac{11\zeta_2}{2}\right)\,\overline{\Sigma}_{V}^{(1,-1)}(0)\, .
 \label{tSigma2bcVxi(2,0)} 
\end{flalign}
\end{subequations}
We are now in a position to use (\ref{Z+Pr vs Sigma 2loop}) and first focus on the renormalization constant. In component form, we obtain:
\begin{subequations}
\label{Z vs Sigma 2loop xi}
\begin{flalign}
	Z_{\psi}^{(2,-2)}(\xi) &=  Z_{\psi}^{(2,-2)}(0) - \xi\,Z_{\psi}^{(1,-1)}(0) + \frac{\xi^2}{2} \, ,
	\label{Z(2,-2) xi} \\
	Z_{\psi}^{(2,-1)}(\xi) &=  Z_{\psi}^{(2,-1)}(0) \!+\! L_p\,\bigg( ( Z_{\psi}^{(1,-1)}(0))^2 - 2 Z_{\psi}^{(2,-2)}(0) \bigg) \nonumber\\
	&=  Z_{\psi}^{(2,-1)}(0) \, ,
        \label{Z(2,-1) xi}
\end{flalign}
\end{subequations}
where
\begin{subequations}
\label{Z vs Sigma 2loop xi=0}
\begin{flalign}
        &Z_{\psi}^{(2,-2)}(0) =  \widetilde{\Sigma}_{bc V}^{(2,-2)}(0) \, ,
        \label{Z(2,-2) xi=0} \\
        &Z_{\psi}^{(2,-1)}(0) =  -4N_F\zeta_2 + \widetilde{\Sigma}_{bc V}^{(2,-1)}(\xi) - \overline{\Sigma}_{V}^{(1,-1)}(0)\,\overline{\Sigma}_{V}^{(1,0)}(0)\, .
        \label{Z(2,-1) xi=0}
\end{flalign}
\end{subequations}
In Eq.~(\ref{Z(2,-1) xi}) we have used a \textit{renormalization constraint} arising from the \textit{finiteness} of the fermion anomalous dimension in the limit $\ep \ra 0$ 
whereby the coefficients $Z_\psi^{(l,-k)}$ for $l>1$ and $k=2,\cdots,l$ may be expressed in terms of coefficients of lower $l$ and $k$. At two-loop, there is only one constraint:
$Z_{\psi}^{(2,-2)}(\xi) = ( Z_{\psi}^{(1,-1)}(\xi) )^2 /2$ which, when applied to (\ref{Z(2,-1) xi}), insures that the renormalization constant 
does not depend on $L_p$. This agrees with the fact that renormalization constants should not depend on masses and external momenta in the $\rm{MS}$ 
scheme~\cite{Vladimirov:1979zm}.

We may proceed in a similar way with the 2-loop renormalized fermion propagator. To leading order in the $\ep$-expansion, it has the form:
\begin{subequations}
\label{P vs Sigma 2loop xi}
\begin{flalign}
	&P_{2r}(p, \xi) =  P_{2r}(p, 0) + \xi\,(L_p - 2)\,\overline{\Sigma}_{V}^{(1,0)}(0) -\nonumber\\
	&\hspace{2cm} - \xi\,(4 - 2\zeta_2 - 2L_p + L_p^2)\,\overline{\Sigma}_{V}^{(1,-1)}(0) + 
	\nonum \\
	&\hspace{2cm} + \xi^2\,\left( 4 - \zeta_2 - 2L_p + \frac{L_p^2}{2}\right)   \, ,
        \label{P2r xi} \\
        &P_{2r}(p, 0) = \widetilde{\Sigma}_{V}^{(2,0)}(0) - 2 L_p\,\widetilde{\Sigma}_{V}^{(2,-1)}(0)  + 2 L_p^2\,\widetilde{\Sigma}_{V}^{(2,-2)}(0) -\nonumber\\
        &\hspace{1.48cm}-
	\overline{\Sigma}_{V}^{(1,-1)}(0)\,\overline{\Sigma}_{V}^{(1,1)}(0) +  
	\nonum \\
	&\hspace{1.48cm} + L_p\,\overline{\Sigma}_{V}^{(1,-1)}(0)\,\overline{\Sigma}_{V}^{(1,0)}(0) - \frac{L_p^2}{2}\,\bigg( \overline{\Sigma}_{V}^{(1,-1)}(0) \bigg)^2\,.
        \label{P2r xi=0}  
\end{flalign}
\end{subequations}

We are now in a position to compare the above derived LKF expressions with the exact results presented in Sec.~\ref{Sec:LKF-renormalization:PT}. 
At one-loop, we find a perfect agreement for the terms proportional to $\xi$ between Eqs.~(\ref{Z+Pr vs Sigma 1loop xi}) and (\ref{Z+Pr:1loop-expression}). 
At two-loop, we also find a perfect agreement for the terms proportional to $\xi^2$ between 
Eqs.~(\ref{Z vs Sigma 2loop xi}) and (\ref{Zpsi:2loop-expression}) on the one hand and between 
Eqs.~(\ref{P2r xi}) and (\ref{Pr:2loop-exppression}) on the other hand. 
These results are in accordance with the fact that \textit{at $l$-loops}, the LKF transformation allows to \textit{fix exactly all terms proportional to 
$\xi^l$}. 

Moreover, by extracting the values of the coefficients $\overline{\Sigma}_{V}^{(1,j)}(0)$ from Eqs.~(\ref{Sigma1a}) and (\ref{Sigma2.4a}) and substituting them 
in Eqs.~(\ref{Z+Pr vs Sigma 1loop xi=0}), (\ref{Z vs Sigma 2loop xi=0}) and (\ref{P2r xi=0}), we immediately recover from
(\ref{Z+Pr vs Sigma 1loop xi}), (\ref{Z vs Sigma 2loop xi}) and (\ref{P2r xi}) the full results of Eqs.~(\ref{Z+Pr:1loop-expression}) and (\ref{Z+Pr:2loop-expression}). 

Finally, we note the remarkable fact that Eq.~(\ref{Z(2,-1) xi}) does not depend on the gauge fixing parameter. From Eq.~(\ref{gammapsi:gen}), this implies that the 2-loop fermion anomalous dimension is \textit{gauge invariant} and is  
in agreement with (\ref{gammapsi:express}). Actually, we may extend such a remark to 3-loops though no exact result is available yet at this order.
All calculations done this yields (in the $\overline{\rm MS}$ scheme):~\footnote{Notice that, while 2-loop results are scheme independent, it is \textit{not} the case
of the 3-loop ones which would take a different form in an alternate scheme choice.}
\begin{subequations}
\label{Z vs Sigma 3loop xi}
\begin{flalign}
        &Z_{\psi}^{(3,-3)}(\xi) = \frac{(1 - 3\xi)^3}{162} \, ,
        \label{Z(3,-3) xi} \\
	&Z_{\psi}^{(3,-2)}(\xi) = - \frac{4\,(1 - 3\xi)}{81}\,\big( 27\,N_F\,\zeta_2 + 4 \big) \, ,  
        \label{Z(3,-2) xi} \\
	&Z_{\psi}^{(3,-1)}(\xi) = \overline{\Sigma}_{V}^{(3,-1)}(0) +\nonumber\\
	&\hspace{2cm} + \frac{\zeta_2}{9}\,\left( 8 N_F\, (5 + 6 \ln 2) - \frac{245}{12} \right) + \frac{1076}{243}\, ,  
        \label{Z(3,-1) xi}
\end{flalign}
\end{subequations}
where the first two terms are easily derived from renormalization constraints~\footnote{Some of these constraints at low orders read:
$Z_{\psi}^{(3,-3)} = Z_{\psi}^{(2,-2)}\,Z_{\psi}^{(1,-1)} - \big( Z_{\psi}^{(1,-1)}\big)^3/3$ and
$Z_{\psi}^{(3,-2)} = Z_{\psi}^{(2,-1)}\,Z_{\psi}^{(1,-1)}$.} while in the third term the Landau gauge coefficient 
$\overline{\Sigma}_{V}^{(3,-1)}(0)$ is not known at the time of writing. Nevertheless, Eq.~(\ref{Z(3,-1) xi}) is 
clearly \textit{gauge invariant} and so is the 3-loop fermion anomalous dimension.

\subsection{Gauge dependence of $\gamma_\psi$}
\label{Sec:LKF-renormalization:gamma}

In the last subsection, the LKF transformation revealed that both the 2-loop and 3-loop fermion anomalous dimensions are
gauge invariant in reduced QED. We will now show that this gauge invariance extends to all higher orders,  see 
Refs.~[\onlinecite{Grozin:2010wa,Kissler:2018lnn}], for similar proofs in the case of QED$_4$.

We proceed in $x$-space starting from the unrenormalized fermion propagator of Eq.~(\ref{LKFN}). 
Similar to the $p$-space case, it is conveniently factored as:
\be
S_F(x, \xi) = Z_\psi(A,\xi)\,S_{Fr}(x,\xi)\, .
\ee
Taking the logarithm of Eq.~(\ref{LKFN})  with $D(x)$ given by Eq.~(\ref{DxN}) and identifying powers of $1/\ep$ straightforwardly yields:
\bea
\log Z_\psi(A,\xi) = \log Z_\psi(A,\eta) - \frac{A\,\Delta}{\ep}\, ,
\label{logZpsi:gen}
\eea
which simply  translates an exponentiation of the gauge-dependence at the level of the renormalization constant.
At this point, let us recall that:
\be
\gamma_\psi(A,\xi) = - \beta(A)\,\frac{\partial \log Z_\psi(A,\xi) }{\partial A}
- \xi\,\gamma(A)\,\frac{\partial \log Z_\psi(A,\xi) }{\partial \xi}\, ,
\label{gammapsi:gen1}
\ee
where $\beta(A)$ is the beta function and $\gamma(A)$ is the gauge-field anomalous dimension. The latter can be expressed as:
\be
\beta(A) = - 2\ep\,A + \sum_{l=1}^{\infty} \beta_l\,A^{l+1}, \quad \gamma(A) = \sum_{l=1}^{\infty} \gamma_{l}\,A^{l}\, ,
\label{beta+gammaA:gen}
\ee
where the coefficients satisfy: $\beta_l = - \gamma_{l}$ (actually, they even vanish in the case of RQED$_{4,3}$). 
Substituting Eq.~(\ref{logZpsi:gen}) in (\ref{gammapsi:gen1}) and using (\ref{beta+gammaA:gen}), yields:
\be
\gamma_\psi(A,\xi) = \gamma_\psi(A,\eta) - 2\,A\,\Delta\, ,
\label{AD-1lxi}
\ee
showing that all the gauge dependence is contained in the one-loop contribution while all higher order
corrections are indeed gauge-invariant.

\section{Summary and Conclusion}
\label{Sec:Conclusion}

In this paper, we have studied the gauge-covariance of the fermion propagator of reduced QED with the help of the 
LKF transformation in dimensional regularization. The $x$-space transformation has been derived in the general case of
QED$_{d_\gamma,d_{\rm e}}$ and its structure, Eq.~(\ref{DxN}), was found to be similar to QED$_4$. Focusing on the 
odd-dimensional case, $d_{\rm e}=3$ (together with $d_\gamma=4-2\ep$), we have then derived the $p$-space LKF transformation
in the form of a series representation for the coefficients of the loop expansion of the propagator in the $\overline{\textrm{MS}}$-scheme, 
Eq.~(\ref{axi.2}) (see also App.~\ref{App:Scales} for other choices of scales and App.~\ref{App:ScalarRQED} for an analogous expression in the case of reduced scalar QED). 
The series has been expressed in terms of a \emph{uniform transcendental factor} $\Phi(m,l,\ep)$, Eq.~(\ref{Phi}). The $\zeta$-structure
of the latter (see Eq.~(\ref{p.s}) and discussion below it) is transcendentally more complicated than in the 
four-dimensional case~\cite{Kotikov:2019bqo} as expected from an odd-dimensional theory~\cite{Broadhurst:1996yc}.  
We have then performed a two-loop expansion of the transformation for the bare fermion propagator, Eqs.~(\ref{OSigma.1+2}),
and also for the renormalization constant and renormalized propagator, Eqs.~(\ref{Z+Pr vs Sigma 1loop xi}), (\ref{Z vs Sigma 2loop xi}) 
and (\ref{P vs Sigma 2loop xi}). 
Starting from the Landau gauge ($\xi = 0$) to a general $\xi$-gauge, 
all these weak-coupling expansions were found to fully agree with previously known exact perturbative results up to the 2-loop order. 
In particular, we have checked that the LKF predicted coefficients of the form $(A\xi)^l$ match with the perturbative results. 
Additionally, we have presented a proof of the purely one-loop gauge dependence of the fermion anomalous dimension in reduced 
QED, Eq.~(\ref{AD-1lxi}). In conclusion, our analysis and in particular our all order series representations, 
Eq.~(\ref{axi.2}) and equivalent ones in App.~\ref{App:Scales}, can of course be used beyond the present 2-loop accuracy of 
perturbative results. They should provide some stringent constraints on future higher order calculations in reduced QED.

\acknowledgments

The work of A.~J.\ is supported by the ILP LABEX (under reference ANR-10-LABX-63) through French state funds managed by the ANR within the Investissements d'Avenir programme under reference ANR-11-IDEX-0004-02.

\appendix

\section{Other choices of scale}
\label{App:Scales}

The calculations in the main text were all performed in the $\overline{\rm MS}$ scheme on the basis of Eqs.~(\ref{axi.2}) and (\ref{OPhi}). 
Following [\onlinecite{Kotikov:2019bqo}], in this Appendix we present three other choices of scale which may be more convenient for future higher 
loop computations: the $g$-scale, the reduced $g$- (or $g_R$-) scale and the MV-scale. We therefore define:
\be
a_m(\xi) = a_m(\eta)
\sum_{l=0}^{\infty} \,
\overline{\Phi}_p(m,l,\ep)
\, \frac{(\Delta \, A)^l}{(-\ep)^l l!} \, {\left(\frac{\mu_p^2}{4p^2}\right)}^{l\ep} \, ,
\label{axi.2.p}
\ee
with
\begin{equation}
\overline{\Phi}_p(m,l,\ep) = \frac{(1-2(m+1)\ep)}{(1-2(m+l+1)\ep)} \, \Phi_p(m,l,\ep) \, ,
\label{OPhi.p}
\end{equation}
where $p = g,~g_R,$ MV and the following subsections will focus on the computation of $\Phi_p(m,l,\ep)$ for these scales. 
In the four-dimensional case, these scales are particularly efficient as they allow a complete subtraction of both
the Euler constant $\gamma_E$ and the $\zeta_2$-value [\onlinecite{Vladimirov:1979zm,Chetyrkin:1980pr}] see also
[\onlinecite{Kotikov:2019bqo}] for a recent application to QED$_4$. 
As will be shown below, in the present $(d_\gamma,d_e) = (4,3)$ case, only the Euler constant is completely subtracted and one cannot avoid the proliferation of 
$\zeta_2$ (as well as $\ln 2$) in accordance with the \textit{greater transcendental complexity} of odd dimensional field theories with respect to even ones [\onlinecite{Broadhurst:1996yc}].

\subsection{$g$-scale}

First, let us consider the so called $G$-scale [\onlinecite{Chetyrkin:1980pr}] which subtracts the coefficient in factor of the singularity $1/\ep$ in the one-loop scalar p-type integral $G(1,1)$. Recalling that:
\be
G(\al,\beta) = \frac{a(\al)\,a(\beta)}{a(\al+\beta-d/2)}\, ,
\ee
where $a(\al)$ was defined in Eq.~(\ref{SFpxN}), the $G$-scale amounts to the following substitution:
\be
\mu_G^{2\ep}=\tilde{\mu}^{2\ep} \,\ep\,G(1,1) =\tilde{\mu}^{2\ep} \, \frac{\Gamma^2(1-\ep)\Gamma(1+\ep)}{\Gamma(2-2\ep)}\, .
\label{G-scale}
\ee
Following [\onlinecite{Broadhurst:1999xk}], a slight modification of this scale that was referred to as the $g$-scale in [\onlinecite{Kotikov:2019bqo}]
subtracts an additional factor $1/(1-2\ep)$ from the one-loop result, \ie,
\be
\mu_g^{2\ep}=\tilde{\mu}^{2\ep} \, \frac{\Gamma^2(1-\ep)\Gamma(1+\ep)}{\Gamma(1-2\ep)}\, .
\label{g-scale}
\ee
With this choice of scale, we have:
\be
\Phi_g(m,l,\ep) = \Phi(m,l,\ep) \times \, e^{-l \gamma_E \ep}\,\frac{\Gamma^l(1-2\ep)}{\Gamma^{2l}(1 - \ep)\,\Gamma^l(1+\ep)}\, ,
\ee
where $\Phi(m,l,\ep)$ is given by Eq.~(\ref{Phi}).
Hence we obtain
\be
\Phi_g(m,l,\ep)= \exp \Big[ \sum_{s=2}^{\infty}\,\eta_s \, p_s^{(g)}(m,l) \, \ep^s \Bigr]\, ,
\label{Phi.g}
\ee
where (for $s \geq 1$)
\begin{align}
p_s^{(g)}(m,l) &= (2^s-1) \, \Bigl\{l + (m+1)^s-(m+l+1)^s \Bigr\} + \nonumber\\
&\qquad{} + (-1)^s \Bigl\{(m+l)^s - m^s - l \Bigr\}\, ,
\label{p.s.g}
\end{align}
and
\be
p_1^{(g)}(m,l)=0,~~ p_{s>1}^{(g)}(m,l) = l\,\Big(2^s - 2 - (-1)^s \Bigr) + p_s(m,l)\, ,
\label{p.2.g}
\ee
\ie, the Euler constant $\gamma_E$ is completely subtracted as in (\ref{p.s}).

\subsection{Reduced $g$-scale}

For reduced QED, it is more natural to consider the $G$ function $G(1, 1 - \varepsilon_{\rm e})$ to define a scale, because it corresponds to the one-loop master integral entering the fermion self-energy in this theory. Thus we write instead
\begin{align}\label{reduced g-scale}
	\mu_{g_R}^{2\ep} &= \tilde{\mu}^{2 \varepsilon} \varepsilon (1 - 2 \varepsilon) G(1, 1 - \varepsilon_{\rm e}) \nonumber \\
	&= \tilde{\mu}^{2\varepsilon} \frac{\Gamma(1 - \varepsilon_{\rm e} -\varepsilon)\Gamma(1 - \varepsilon) \Gamma(1+\varepsilon)}{\Gamma(1-\varepsilon_{\rm e}-2\varepsilon)}\, ,
\end{align}
which for $\textrm{RQED}_{4,3}$ becomes
\begin{align}
\mu^{2\varepsilon}_{g_R} = \frac{\tilde{\mu}^{2\varepsilon}}{4^\varepsilon} \frac{\Gamma^2(1-2\varepsilon)\Gamma(1+\varepsilon)}{\Gamma(1-4\varepsilon)}\, .
\end{align}
This leads to a $\Phi_{g_R}$ function 
\be
\Phi_{g_R}(m,l,\ep) = \Phi(m,l,\varepsilon) \times e^{-l \gamma_E \varepsilon} \, 4^{l \varepsilon}  
\frac{\Gamma^l(1 - 4\varepsilon)}{\Gamma^{2l}(1 - 2 \varepsilon) \Gamma^l(1 + \varepsilon)}\, .
\ee
Hence:
\be\label{Phi.gr}
\Phi_{g_R}(m,l,\varepsilon) = 4^{l\varepsilon}\exp\left[\sum_{s=2}^\infty p_s^{(g_R)}(m,l) \eta_s \varepsilon^s\right]\, , 
\ee
where
\begin{align}\label{p.s.g.r}
p_{s}^{(g_R)}(m,l) &= (2^s -1) \Bigl\{(m+1)^s - (m+l+1)^s\Bigr\} + \nonumber\\
&\quad{} + (2^s-1)^2 l + (-1)^s\Bigl\{(m+l)^s - m^s -l\Bigr\},
\end{align}
such that
\begin{align}
p_1^{g_R}(m,l) = 0, \qquad p_{s > 1}^{g_R}(m,l) &= l\left(2^{2s} - 2^{s+1} - (-1)^s \right) +\nonumber\\
&\qquad{} + p_s(m,l) \, ,
\label{p.1+2.g reduced}
\end{align}
\ie, the Euler constant $\gamma_E$ is completely subtracted as in (\ref{p.s}).

\subsection{MV-scale}

Yet another convenient choice of scale is the minimal Vladimirov-scale [\onlinecite{Kotikov:2019bqo}] which is defined via the relation
\be
\mu_{\text{MV}}^{2\ep} = \frac{\tilde{\mu}^{2\ep}}{\Gamma(1-\ep)} \, .
\label{MV-scale}
\ee
With this choice of scale, we have:
\be
\Phi_{\rm MV}(m,l,\ep) = \Phi(m,l,\ep) \, e^{-l \gamma_E \ep}\,\Gamma^l(1-\ep)\, ,
\ee
where $\Phi(m,l,\ep)$ is given by Eq.~(\ref{Phi}).
Hence:
\be
\Phi_{\rm MV}(m,l,\ep)= \exp \Big[ \sum_{s=2}^{\infty}\,\eta_s \, p_s^{({\rm MV})}(m,l) \, \ep^s \Bigr]\, ,
\label{Phi.MV}
\ee
where (for $s \geq 1$)
\begin{align}
p_s^{({\rm MV})}(m,l)&= (2^s-1) \, \Bigl\{(m+1)^s-(m+l+1)^s \Bigr\} + 2l + \nonumber\\
&\qquad{} + (-1)^s \Bigl\{(m+l)^s - m^s  \Bigr\} \, ,
\label{p.s.MV}
\end{align}
and
\be
p_1^{({\rm MV})}(m,l)=0,~~ p_{s>1}^{({\rm MV})}(m,l) = l + p_s(m,l) \, ,
\label{p.2.MV}
\ee
\ie, the Euler constant $\gamma_E$ is completely subtracted as in (\ref{p.s}).

 We can see that in all cases considered, \ie, in $g$, reduced $g$ and $MV$-scales, we cannot put the values $p_{s=2}^{({\rm i})}(m,l)$ to be zero as it was before
  in the case of the spinor and scalar QED (see Ref. [\onlinecite{Kotikov:2019bqo}]). Indeed,  $p_{s=2}(m,l)$ has the exact $m$-dependence as  is shown in (\ref{p.2}). So,
  contrary to the QED$_4$ case, the coefficients of $\ep$-expansion in the case of QED$_{4,3}$ contain exactly these $\zeta_2$ values.
  However, let us note that $p_{s=2}^{({\rm MV})}(m,l)=p_{s=2}^{({\rm g})}(m,l)$ as it was in the QED$_4$ case.

\section{Reduced scalar QED$_{4,3}$}
\label{App:ScalarRQED}

For completeness, we shall consider here the case of reduced scalar (spin-$0$) QED which is similar
to reduced spinor QED that has been considered throughout the rest of the paper.

These (massless) models have the Lagrangian (in Minkowski space)
\begin{align}
\mathcal{L} =  \lvert D_\mu \phi\rvert^2 -\frac{1}{4} F_{\mu\nu} F^{\mu\nu} - \frac{1}{2\xi} (\partial_\mu A^\mu)^2 + \frac{\lambda}{4!}(\lvert \phi \rvert^2)^2
\end{align}
to be integrated over the appropriate volume element for the theory under consideration. As before, we only focus on the gauge covariance of the scalar propagator.

The general expression of a scalar propagator, $S_C(p,\eta)$, of external momentum $p$ and gauge fixing parameter $\eta$
reads:
\be
S_C(p, \eta) = \frac{1}{p^2} \, \sum_{m=0}^{\infty} a^c_m(\eta)\, A^m \,{\left(\frac{\tilde{\mu}^2}{p^2}\right)}^{m\ep} \, ,
\label{Sc.eta}
\ee
where $a^c_m(\eta)$ are the coefficients of the loop expansion of the propagator and
$\tilde{\mu}$ is the renormalization scale (\ref{Aem}).

Proceeding in a way similar to the spinor case, the scalar propagator in another gauge $\xi$
is obtained from the following LKF transformation:
\be
S_C(p,\xi) = \frac{1}{p^2} \, \sum_{m=0}^{\infty} a^c_m(\xi)\, A^m \,{\left(\frac{\tilde{\mu}^2}{p^2}\right)}^{m\ep} \, ,
\label{Sc.xi}
\ee
where
\begin{flalign}
	a^c_m(\xi) &= a^c_m(\eta)\, \frac{\Gamma(1-\ep_{\rm e} -(m+1)\ep)}{\Gamma(1+m\ep)} \, \times
\nonum \\
&\times
\, \sum_{l=0}^{\infty}
\frac{\Gamma(1+(m+l)\ep)\,\Gamma^l(1-\ep)}{l!\,\Gamma(1-\ep_{\rm e}-(m+l+1)\ep)} \, \frac{(\Delta \, A)^l}{(-\ep)^l} \, {\left(\frac{\tilde{\mu}^2}{p^2}\right)}^{l\ep}\, ,
\label{axi.c}
\end{flalign}
which is valid for arbitrary $\ep_{\rm e}$. The only difference with respect to the spinorial case is that, in the latter, there is an additional factor of
$(1-\ep_{\rm e} -(m+1)\ep)/(1-\ep_{\rm e}-(m+l+1)\ep)$ as can be seen from
\begin{flalign}
        a_m(\xi) &= a_m(\eta)\, \frac{\Gamma(2-\ep_{\rm e} -(m+1)\ep)}{\Gamma(1+m\ep)} \, \times
\nonum \\
&\times
\, \sum_{l=0}^{\infty}
\frac{\Gamma(1+(m+l)\ep)\,\Gamma^l(1-\ep)}{l!\,\Gamma(2-\ep_{\rm e}-(m+l+1)\ep)} \, \frac{(\Delta \, A)^l}{(-\ep)^l} \, {\left(\frac{\tilde{\mu}^2}{p^2}\right)}^{l\ep}\, ,
\label{axi.F}
\end{flalign}
which simply generalizes (\ref{axi}) to arbitrary $\ep_{\rm e}$.

In the case of scalar QED$_{4,3}$ ($\ep_{\rm e}=1/2$), Eq.~(\ref{axi.c}) can then be written as:
\be
a_m^c(\xi) = a_m^c(\eta)
\sum_{l=0}^{\infty} \,
\Phi_p(m,l,\ep)
\, \frac{(\Delta \, A)^l}{(-\ep)^l l!} \, {\left(\frac{\mu_p^2}{4p^2}\right)}^{l\ep} \, ,
\label{axi.c.2.p}
\ee
where $\Phi_p(m,l,\ep)$ is given by Eq.~(\ref{Phi}) for the $\overline{\rm MS}$ scale and by Eqs.~(\ref{Phi.g}), (\ref{Phi.gr}), and (\ref{Phi.MV})
for the $g$-, $g_R$-, and MV-scales, respectively. So the only difference between (\ref{axi.2.p}) and (\ref{axi.c.2.p}) is in the factor
$(1 - 2(m+1)\ep)/(1-2(m+l+1)\ep)$ which is absent in the scalar QED$_{4,3}$ case.



\begin{thebibliography}{0}




\bibitem{Landau:1955zz} 
  L.~D.~Landau and I.~M.~Khalatnikov,
  Sov.\ Phys.\ JETP {\bf 2}, 69 (1956)
  [Zh.\ Eksp.\ Teor.\ Fiz.\  {\bf 29}, 89 (1955)];
  E.~S.~Fradkin,
  Zh.\ Eksp.\ Teor.\ Fiz.\  {\bf 29}, 258 (1955)
  [Sov.\ Phys.\ JETP {\bf 2}, 361 (1956)].

\bibitem{Johnson:1959zz} 
  K.~Johnson and B.~Zumino,
  Phys.\ Rev.\ Lett.\  {\bf 3}, 351 (1959);
  B.~Zumino,
  J.\ Math.\ Phys.\  {\bf 1}, 1 (1960).

\bibitem{Sonoda:2000kn} 
  S. Okubo, Nuovo Cimento {\bf 15}, 949 (1960);
  I. Bialynicki-Birula, Nuovo Cimento {\bf 17}, 951 (1960);
  H.~Sonoda,
  Phys.\ Lett.\ B {\bf 499}, 253 (2001).
  
\bibitem{Curtis:1990zs} 
  D.~C.~Curtis and M.~R.~Pennington,
  Phys.\ Rev.\ D {\bf 42}, 4165 (1990);
  Z.~h.~Dong, H.~J.~Munczek and C.~D.~Roberts,
  Phys.\ Lett.\ B {\bf 333}, 536 (1994);
  A.~Bashir and M.~R.~Pennington,
  Phys.\ Rev.\ D {\bf 50}, 7679 (1994);
  Phys.\ Rev.\ D {\bf 53}, 4694 (1996);
  A.~Bashir, A.~Kizilersu and M.~R.~Pennington,
  Phys.\ Rev.\ D {\bf 57}, 1242 (1998).

  
  
\bibitem{Burden:1998gr} 
  C.~J.~Burden and P.~C.~Tjiang,
  Phys.\ Rev.\ D {\bf 58}, 085019 (1998);
  A.~Bashir, A.~Kizilersu and M.~R.~Pennington,
  Phys.\ Rev.\ D {\bf 62}, 085002 (2000);
  A.~Bashir and A.~Raya,
  Phys.\ Rev.\ D {\bf 64}, 105001 (2001).

\bibitem{Jia:2016udu}
  S.~Jia and M.~R.~Pennington,
  Phys.\ Lett.\ B {\bf 769}, 146 (2017);
  Phys.\ Rev.\ D {\bf 94}, no. 11, 116004 (2016);
  Phys.\ Rev.\ D {\bf 96}, no. 3, 036021 (2017).

\bibitem{Fernandez-Rangel:2016zac} 
  L.~A.~Fernandez-Rangel, A.~Bashir, L.~X.~Gutierrez-Guerrero and Y.~Concha-Sanchez,
  Phys.\ Rev.\ D {\bf 93}, no. 6, 065022 (2016);
  N.~Ahmadiniaz, A.~Bashir and C.~Schubert,
  Phys.\ Rev.\ D {\bf 93}, no. 4, 045023 (2016).

\bibitem{Kizilersu:2009kg} 
  A.~Kizilersu and M.~R.~Pennington,
  Phys.\ Rev.\ D {\bf 79}, 125020 (2009).


\bibitem{Bashir:2002sp}
  A.~Bashir and A.~Raya,
  Phys.\ Rev.\ D {\bf 66}, 105005 (2002).

\bibitem{Jia:2016wyu} 
  S.~Jia and M.~R.~Pennington,
  Phys.\ Rev.\ D {\bf 95}, no. 7, 076007 (2017).
 
\bibitem{Kotikov:2019bqo} 
  A.~V.~Kotikov and S.~Teber,
  Phys.\ Rev.\ D {\bf 100}, no. 10, 105017 (2019).

\bibitem{Ahmad:2016dsb} 
  A.~Ahmad, J.~J.~Cobos-Mart\'{i}nez, Y.~Concha-S\'{a}nchez and A.~Raya,
  Phys.\ Rev.\ D {\bf 93}, no. 9, 094035 (2016).

\bibitem{DeMeerleer:2018txc} 
  T.~De Meerleer, D.~Dudal, S.~P.~Sorella, P.~Dall'Olio and A.~Bashir,
  Phys.\ Rev.\ D {\bf 97}, no. 7, 074017 (2018).

\bibitem{DeMeerleer:2019kmh} 
  T.~De Meerleer, D.~Dudal, S.~P.~Sorella, P.~Dall'Olio and A.~Bashir,
  arXiv:1911.01907 [hep-th].

\bibitem{Broadhurst:1999xk}
  D.~J.~Broadhurst,
  hep-th/9909185.

\bibitem{Baikov:2010hf}
  P.~A.~Baikov and K.~G.~Chetyrkin,
  Nucl.\ Phys.\ B {\bf 837}, 186 (2010).

\bibitem{Baikov:2018wgs}
  P.~A.~Baikov and K.~G.~Chetyrkin,
  JHEP {\bf 1806}, 141 (2018).

\bibitem{Baikov:2018gap}
  P.~A.~Baikov and K.~G.~Chetyrkin,
  PoS LL {\bf 2018}, 008 (2018).

\bibitem{Baikov:2019zmy} 
  P.~A.~Baikov and K.~G.~Chetyrkin,
  JHEP {\bf 1910}, 190 (2019).

\bibitem{Teber:2012de} 
  S.~Teber,
  Phys.\ Rev.\ D {\bf 86}, 025005 (2012);
  A.~V.~Kotikov and S.~Teber,
  Phys.\ Rev.\ D {\bf 87}, no. 8, 087701 (2013);
  S.~Teber,
  Phys.\ Rev.\ D {\bf 89}, no. 6, 067702 (2014).

\bibitem{Kotikov:2013eha} 
  A.~V.~Kotikov and S.~Teber,
  Phys.\ Rev.\ D {\bf 89}, no. 6, 065038 (2014).

\bibitem{Teber:2018goo} 
  S.~Teber and A.~V.~Kotikov,
  Phys.\ Rev.\ D {\bf 97}, no. 7, 074004 (2018).


\bibitem{PhysRev.71.622}
	P.\ R.~Wallace, 
 Phys.\ Rev.\ {\bf 71} 622 (1947).

\bibitem{Semenoff:1984dq}
G.~W.~Semenoff, 
 Phys.\ Rev.\ Lett.\  {\bf 53} 2449 (1984).

\bibitem{Novoselov:2005kj}
K.~S.~Novoselov, A.~K.~Geim, S.~V.~Morozov, D.~Jiang, M.~I.~Katsnelson, I.~V.~Grigorieva, S.~V.~Dubonos and A.~A.~Firsov, 
 Nature {\bf 438} 197 (2005).

\bibitem{polini2013artificial}
M.~Polini, F.~Guinea, M.~Lewenstein, H.\ C.~Manoharan and V.~Pellegrini, 
  Nature Nanotechnology {\bf 8} 625 (2013).

\bibitem{RevModPhys.82.3045}
M.\ Z.~Hasan and C.\ L.~Kane, 
 Rev.\ Mod.\ Phys.\ {\bf 82} 3045 (2010).

\bibitem{PanCF:2017}
W.~Pan, W.~Kang, K.\ W.~Baldwin, K.\ W.~West, L.\ N.~Pfeiffer and D.\ C.~Tsui, 
 Nature Physics {\bf 13} 1168 (2017).

\bibitem{Kotikov:2016yrn}
  A.~V.~Kotikov and S.~Teber,
  Phys.\ Rev.\ D {\bf 94}, no. 11, 114010 (2016);
  Erratum: [Phys.\ Rev.\ D {\bf 99}, no. 11, 119902 (2019)].

\bibitem{Marston:1989zz}
J.~B.~Marston and I.~Affleck, 
Phys.\ Rev.\ B {\bf 39} 11538 (1989).

\bibitem{Ioffe:1989zz}
L.~B.~Ioffe and A.~I.~Larkin, 
Phys.\ Rev.\ B {\bf 39} 8988 (1989).

\bibitem{Herbut:2002yq} 
  I.~F.~Herbut,
  Phys.\ Rev.\ B {\bf 66}, 094504 (2002).

 \bibitem{Gonzalez:1993uz}
J.~Gonz\'alez, F.~Guinea and M.~A.~H.~Vozmediano, 
 Nucl.\ Phys.\ B {\bf 424} 595 (1994).

\bibitem{Teber:2014ita}
  S.~Teber and A.~V.~Kotikov,
  EPL {\bf 107}, no. 5, 57001 (2014);
  JHEP {\bf 1807}, 082 (2018).

\bibitem{Munoz-Segovia:2019xip} 
D.~Mu\~{n}oz-Segovia and A.~Cortijo,
  arXiv:1910.08081 [cond-mat.str-el].

\bibitem{Teber:2016unz} 
  S.~Teber and A.~V.~Kotikov,
  Theor.\ Math.\ Phys.\  {\bf 190}, no. 3, 446 (2017).
  
\bibitem{Teber:2019kkp} 
  S.~Teber and A.~V.~Kotikov,
  Theor.\ Math.\ Phys.\  {\bf 200}, no. 2, 1222 (2019).


\bibitem{Kotikov:2018wxe} 
  A.~V.~Kotikov and S.~Teber,
  Phys.\ Part.\ Nucl.\  {\bf 50}, no. 1, 1 (2019).


\bibitem{Kotikov:2000pm} 
  A.~V.~Kotikov and L.~N.~Lipatov,
  Nucl.\ Phys.\ B {\bf 582}, 19 (2000).

\bibitem{Kotikov:2002ab} 
  A.~V.~Kotikov and L.~N.~Lipatov,
  Nucl.\ Phys.\ B {\bf 661}, 19 (2003);
  A.~V.~Kotikov, L.~N.~Lipatov, A.~I.~Onishchenko and V.~N.~Velizhanin,
  Phys.\ Lett.\ B {\bf 595}, 521 (2004).

\bibitem{Fleischer:1998nb} 
  J.~Fleischer, A.~V.~Kotikov and O.~L.~Veretin,
  Nucl.\ Phys.\ B {\bf 547}, 343 (1999).

  
\bibitem{Kotikov:2007cy} 
  A.~V.~Kotikov, L.~N.~Lipatov, A.~Rej, M.~Staudacher and V.~N.~Velizhanin,
  J.\ Stat.\ Mech.\  {\bf 0710}, P10003 (2007);
  Z.~Bajnok, R.~A.~Janik and T.~Lukowski,
  Nucl.\ Phys.\ B {\bf 816}, 376 (2009);
  T.~Lukowski, A.~Rej and V.~N.~Velizhanin,
  Nucl.\ Phys.\ B {\bf 831}, 105 (2010);
  C.~Marboe, V.~Velizhanin and D.~Volin,
  JHEP {\bf 1507}, 084 (2015);
  C.~Marboe and V.~Velizhanin,
  JHEP {\bf 1611}, 013 (2016).
  
\bibitem{Dixon:2019uzg} 
  L.~J.~Dixon, I.~Moult and H.~X.~Zhu,
  Phys.\ Rev.\ D {\bf 100}, no. 1, 014009 (2019);
  J.~Broedel, C.~Duhr, F.~Dulat, B.~Penante and L.~Tancredi,
  JHEP {\bf 1905}, 120 (2019).
  
\bibitem{Vladimirov:1979zm}
  A.~A.~Vladimirov,
  Theor.\ Math.\ Phys.\  {\bf 43}, 417 (1980)
  [Teor.\ Mat.\ Fiz.\  {\bf 43}, 210 (1980)].




\bibitem{Grozin:2010wa} 
A.~G.~Grozin,
Phys.\ Lett.\ B {\bf 692}, 161 (2010);
%
 A.~G.~Grozin,
  arXiv:1305.4245 [hep-ph].

\bibitem{Kissler:2018lnn}
  H.~Kissler,
  PoS LL {\bf 2018}, 032 (2018).

\bibitem{Chetyrkin:1980pr}
  K.~G.~Chetyrkin, A.~L.~Kataev and F.~V.~Tkachov,
  Nucl.\ Phys.\ B {\bf 174}, 345 (1980).

\bibitem{Broadhurst:1996yc} 
  D.~J.~Broadhurst and A.~V.~Kotikov,
  Phys.\ Lett.\ B {\bf 441}, 345 (1998).



\end{thebibliography}
\end{document}